\documentclass[12pt,notitlepage,a4paper]{article}

\pdfoutput=1

\usepackage{a4wide}
\usepackage{epsfig}
\usepackage{color,graphicx}
\usepackage{cite}
\usepackage{amssymb,amsmath}

\newcommand{\be}{\begin{equation}}
\newcommand{\ee}{\end{equation}}
\newcommand{\bea}{\begin{eqnarray}}
\newcommand{\eea}{\end{eqnarray}}

\begin{document}

\begin{center}  

\vskip 2cm 

\centerline{\Large {\bf Brane webs in the presence of an $O5^-$-plane}}
\vskip 0.5cm
\centerline{\Large {\bf and $4d$ class S theories of type $D$}}
\vskip 1cm

\renewcommand{\thefootnote}{\fnsymbol{footnote}}

   \centerline{
    {\large \bf Gabi Zafrir${}^{a}$} \footnote{gabizaf@techunix.technion.ac.il}}

\vspace{1cm}
\centerline{{\it ${}^a$ Department of Physics, Technion, Israel Institute of Technology}} \centerline{{\it Haifa, 32000, Israel}}
\vspace{1cm}

\end{center}

\vskip 0.3 cm

\setcounter{footnote}{0}
\renewcommand{\thefootnote}{\arabic{footnote}}   
   
\begin{abstract}

In this article we conjecture a relationship between $5d$ SCFT's, that can be engineered by $5$-brane webs in the presence of an $O5^-$-plane, and $4d$ class S theories of type $D$. The specific relation is that compactification on a circle of the former leads to the latter. We present evidence for this conjecture. One piece of evidence, which is also an interesting application of this, is that it suggests identifications between different class S theories. This can in turn be tested by comparing their central charges. 

\end{abstract}
 
 \newpage
 
\tableofcontents

\section{Introduction}

One of the interesting conclusions from string theory is the existence of supersymmetric conformal field theories in five and six dimensions. These can sometimes be deformed so as to flow to gauge theories in the IR, which are non-renormalizable in these dimensions. Therefore, in a sense these SCFT's provide UV completions to these gauge theories. For example, in $5d$ the SCFT can be perturbed by a mass deformation leading to a low-energy gauge theory where the inverse gauge coupling squared, which has dimension of mass, is identified with the mass deformation\cite{SEI}. Taking the infinite coupling limit recovers the $5d$ SCFT.

It has proven interesting to study the compactification of these SCFT's to four dimensions. The most studied case is undoubtedly the compactification of the $6d$ $(2,0)$ theory on a Riemann surface initially studied in \cite{Gai}. This leads to a large number of interacting four dimensional SCFT's including the so called non-Lagrangian SCFT's such as the Minahan-Nemeschansky $E_6, E_7$ and $E_8$ theories\cite{MN}. These theories are usually referred to as class S theories.

In this article we will be mostly concerned with compactification of $5d$ SCFT's to $4d$. This was studied in \cite{BB} for a specific class of $5d$ SCFT's where the compactification results in a $4d$ class S theory of type $A$. The $5d$ SCFT's considered there can be engineered in string theory by brane webs\cite{HA,AHK}. In the simplest case, where the $5d$ SCFT is described by a $5$-brane junction of $(1,0)$, $(0,1)$ and $(1,1)$ $5$-branes, the $4d$ class S theory is a compactification of an $A$ type $(2,0)$ theory on a three punctured sphere. Thus, in this case the resulting $4d$ theory is an isolated SCFT which in general has no Lagrangian description. One can then study the $5d$ SCFT which can lead to interesting consequences also for the $4d$ SCFT, see \cite{BMPTY,HKT,BZ,HTY}. Also note that compactifying a $5d$ SCFT does not always result in a $4d$ SCFT, see \cite{OSTY2} for an example.   

There are a large number of other $5d$ SCFT's, some of which can also be engineered by generalizations of brane webs through the addition of orientifold planes\cite{KB,BZ1,Zaf3,HKLTY1}. In this article we conjecture that compactifying a class of $5d$ SCFT's that can be engineered by brane webs in the presence of an $O5^-$-plane leads to $4d$ type $D$ class S theories. This can be motivated by studying the dimension of the Higgs branch, and the global symmetries manifest in the web description, and matching them against the properties of class S theories. 

The major piece of evidence we provide, which is also an interesting application of this relation, is by testing equivalences between different class S theories. Many of the $5d$ SCFT's we consider can be mass deformed to a $5d$ gauge theory. In some cases this gauge theory can also be realized by a different brane realization, either with the $O5^-$-plane, or without it which leads to a web of the type considered in \cite{BB}. In both descriptions the SCFT is realized when all mass parameters vanish. Thus, from $5d$ reasoning we conclude that these are different string theory realizations of the same $5d$ SCFT. On one hand compactifying this $5d$ SCFT should lead to one $4d$ SCFT. On the other hand, as the string theory realizations are different, we get seemingly different class S theories. Consistency now necessitates that these class S theories are in fact identical. This can be checked by comparing the central charges of these class S theories. We indeed find that they are equal in all cases checked. 

The structure of this article is as follows. Section 2 consists of a summary of properties of $D$ type class S theories and brane webs in the presence of an $O5^-$-plane that play a role in this article. In section 3 we present our conjecture and provide supporting evidence for it. We end with some conclusions. The appendix discusses the suspected creation of free hypermultiplets accompanying certain $7$-brane manipulations.  


 


\section{Preliminaries}

We begin by reviewing several aspects of class S technology and brane webs in the presence of $O5$-planes that play an important role in the proceeding discussion. 

\subsection{Class S technology}


In this article we will be particularly interested in $4d$ class S SCFT's given by compactifying the type $D$ $(2,0)$ theory on a Riemann sphere with punctures. There are known methods to compute various quantities of interest for this class of theories, and in this subsection we shall summarize the ones used in this article. All of these, and much more, can be found in \cite{Tachi1,CD1}.

Like their $A$ type cousins, punctures of $D$ type class S theories are labeled by Young diagrams. For a $D_N$ theory, this is given by a Young diagram with $2N$ boxes, subject to the constraint that columns made of an even number of boxes must repeat with even multiplicity. Furthermore if all the columns are made of an even number of boxes then there are actually two different punctures associated with this Young diagram. These are usually called very-even partitions, and the two punctures are usually distinguished by the color of the diagram, red and blue being the common choice. 

Each puncture has a global symmetry associated to it, where each group of $n_h$ columns with the some height $h$ contribute a $USp(n_h)$ factor if $h$ is even, or an $SO(n_h)$ factor if $h$ is odd. Note that here the constraint that columns made of an even number of boxes must repeat with even multiplicity is important. The contributions from all the punctures then give part of the global symmetry of the SCFT, which may be further enhanced to a larger global symmetry. In some cases we will also want to determine the full global symmetry. In those cases we use the $4d$ superconformal index.  

 Since conserved currents are BPS operators they contribute to the index, and so knowledge of the index allows us to determine the global symmetry of the theory. In practice we do not need the full superconformal index, just the first few terms in a reduced form of the index called the Hall-Littlewood index\cite{GRRY}. an expression for the $4d$ superconformal index for class S theories was conjectured in \cite{GRRY,GR,GRR}, and further studied for the $D$ type case in \cite{LPR}. Occasionally a class S theory may contain free hypermultiplets in addition to, or even without, a strongly interacting part. The $4d$ index then provides an excellent tool for discovering if such a thing occurs, and to find the number of such free hypers. A good review for these applications of the index of class S theories can be found in \cite{CDT}.

We will also need to evaluate various properties of the SCFT, notably, the spectrum and dimension of Coulomb branch operators, dimension of the Higgs branch, and central charges of the various global symmetries. Their evaluation from the Riemann surface is known and given in \cite{CD1}. One quantity that will play an important role is the dimension of the Higgs branch for a SCFT arising from the compactification on a three punctured sphere. The Higgs branch dimension is then given by:

\be
d_H = N + f_1 + f_2 + f_3 \label{DoHB}
\ee 
where the compactified theory is $D_N$, and the $f_i$'s are the contributions of the three punctures. These can be evaluated directly from the Young diagram, and are given by:

\be
f = \frac{1}{4} \sum_j r^2_j - \frac{1}{2} \sum_{\text{j odd}} r_j \label{HBfP}
\ee
where $r_j$ is the length of the $j$'th row, and the first sum is over all rows while the second is only over the odd numbered rows ($j=1,3,5...$).

\subsection{Brane webs in the presence of $O5$-planes}

In this subsection we shall summarize some important results regarding brane webs in the presence of $O5$-planes. Such systems were first introduced in \cite{KB}, and further studied in \cite{Zaf3,HKLTY1}. The particular system we are interested in consists of a group of D$5$-branes, an $O5^-$-plane parallel to the D$5$-branes and several NS$5$-branes stuck on the orientifold $5$-plane. Such a stuck NS$5$-brane leads to a change in the orientifold type: an $O5^-$-plane changes to an $O5^+$-plane and vice versa. We also take the number of such branes to be even so that the asymptotic orientifold $5$-plane on both sides is an $O5^-$-plane. We can further add $7$-branes on which the $5$-branes can end.

 The gauge symmetry living on $N$ D$5$-branes parallel to an $O5^-$-plane ($O5^+$-plane) is $SO(2N)$ ($USp(2N)$). Thus, the resulting $5d$ SCFT has mass deformations leading to a $5d$ quiver with alternating $SO$ and $USp$ gauge groups. The matter content includes half-bifundamentals between each $USp/SO$ pair, fundamental hypermultiplets for the $USp$ groups and vector matter for the $SO$ groups. In some cases we can also incorporate spinor matter for some $SO$ groups. We refer the reader to \cite{Zaf3} for the details. 

Besides the $USp/SO$ quiver, these SCFT's have an additional gauge theory description. In the brane web this can be seen by performing S-duality on the configuration where the $O5^-$-plane becomes a perturbative orbifold. Specifically, an $O5^-$-plane with a full D$5$-brane on top of it is S-dual to an object called an $ON^0$ plane, which is a $\mathbb{C}^2/I_4 (-1)^{F_L}$ orbifold\cite{HZ,HK}. Studying the gauge theory living on the $5$-branes in this background reveals that the $5d$ SCFT also possess mass deformations leading now to a $5d$ quiver of $SU$ groups in the shape of a $D$ type Dynkin diagram.

 We will also want to understand the global symmetry of the SCFT. This can be determined by considering the gauge symmetry living on the $7$-branes, which is a global symmetry from the point of view of the $5$-branes. For the system we will be considering, there are three contributions given by the D$7$-branes on each side of the orientifold as well as the $(0,1)$ $7$-branes. At the fixed point the D$7$-branes, on each side of the orientifold, merge on the $O5^-$-plane while the $(0,1)$ $7$-branes merge outside the $O5^-$-plane. 

We want to consider the classical gauge symmetry on the D$7$-branes, which may also have an arbitrary number of D$5$-branes ending on them. To do this we envision moving them through several NS$5$-branes. Each such transition removes a D$5$-brane via the Hanany-Witten effect, so as to arrive at a group of D$7$-branes with no D$5$-branes ending on them. Once this configuration is reached, the D$7$-branes sit on top of an $O5^+$-plane ($O5^-$-plane) if, in the initial configuration, the number of D$5$-branes ending on the D$7$-branes is odd (even). The gauge symmetry living on a group of $k$ stuck D$7$-branes intersecting an $O5^-$-plane ($O5^+$-plane) is $USp(k)$ ($SO(k)$) where $k$ must be even for the $O5^-$ case. 

For the $(0,1)$ $7$-branes each group of $n$ such $7$-branes with the same number of NS$5$-branes ending on them contributes a $U(n)$ global symmetry. Taking the direct product of all three contributions then gives the classical global symmetry. This may be further enhanced to a larger symmetry group. 

\begin{figure}
\center
\includegraphics[width=0.5\textwidth]{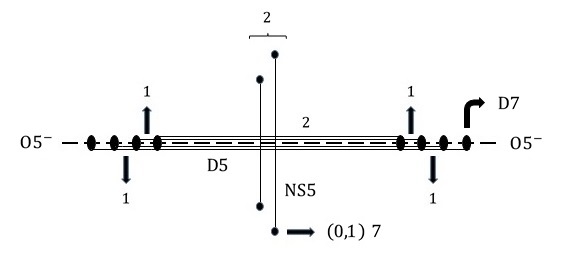} 
\caption{An example for the Higgs branch of a $5d$ SCFT of the form considered in this article. The total Higgs branch dimension in this example is $8$, which we separate to four contributions which in this case happen to be equal. First we can pairwise break and separate the D$5$-branes along the D$7$-branes on one side of the orientifold. In this case there are two such directions for each side of the $O5$-plane, where there is an arrow representing the pair. There are also two directions given by pairwise breaking and separating the four D$5$-branes stretching between the two groups of D$7$-branes on either side of the $O5$-plane. Finally we can pairwise break and separate the NS$5$-branes along the $(0,1)$ $7$-branes. In this case there are two directions, one given by breaking the longer part of the right NS$5$-branes while the other by separating the remaining part of the NS$5$-branes from the $O5$-plane.}
\label{Rd7}
\end{figure}

We also wish to determine the dimension of the Higgs branch of the SCFT. This can be done from the web by examining the number of motions of the $5$-branes parallel to the $7$-branes. For the system we consider, these can be broken into four parts. First we can separate the D$7$-branes across the $O5^-$-plane, and pairwise separate D$5$-branes, stretched between neighboring D$7$-branes, from the orientifold (see the discussion in \cite{Zaf3}). Doing this on either side of the orientifold comprises parts one and two, and doing so on the central $5$-branes comprises part three. Finally we can break  the NS$5$-branes on the $(0,1)$ $7$-branes and pairwise separate them from the $O5^-$-plane. This gives the fourth part. See figure \ref{Rd7} for an example.
 
\section{Statement of the conjecture}

In this section we present our conjecture that compactifying a class of $5d$ SCFT's on a circle leads to $4d$ class S SCFT's given by compactifying the type $D$ $(2,0)$ theory on a Riemann sphere with punctures. We start with the cases related to compactification on a three punctured sphere, and then move on to the general case.

\subsection{Isolated $4d$ SCFT's}

 We start by presenting the class of $5d$ SCFT's we consider. These can be engineered by a web with an $O5$-plane, and a system of $5$-branes ending on $(1,0)$ or $(0,1)$ $7$-branes only. More specifically, asymptotically the orientifold plane is $O5^-$ in both directions, and there are a total of $2N$ D$5$-branes (in the covering space) ending on the $(1,0)$ $7$-branes whose number may be less than $2N$. The number of $(0,1)$ $7$-branes is at most two. If there is only one all asymptotic NS$5$-branes end on it, while if there are two then one NS$5$-brane ends on one of the $(0,1)$ $7$-branes, and the rest end on the other. 

We claim that compactifying the $5d$ SCFT on a circle of radius $R_5$ and taking the $R_5\rightarrow 0$ limit, without mass deformations, leads to a $4d$ class S theory given by compactifying the type $D_N$ $(2,0)$ theory on a Riemann sphere with three punctures. The three punctures are associated with the $7$-branes, and the $5$-branes ending on them, where one puncture is associated with the $(0,1)$ $7$-branes and two with the $(1,0)$ $7$-branes on either side of the orientifold.

\begin{figure}
\center
\includegraphics[width=0.8\textwidth]{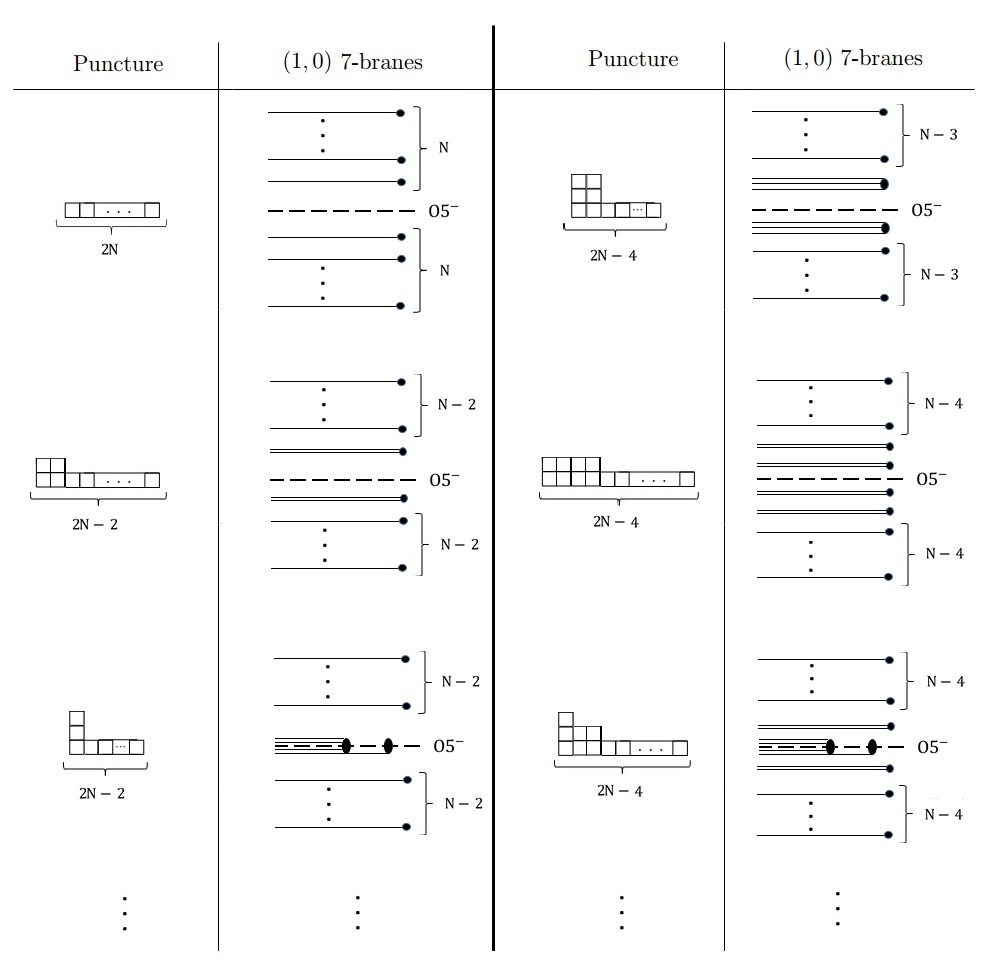} 
\caption{The mapping between the allocation of D$5$-branes on the $(1,0)$ $7$-branes and the associated puncture. In all cases the $6d$ theory is the $(2,0)$ theory of type $D_{N}$ where $N$ is determined as explained in the text.}
\label{Rd0}
\end{figure}

The punctures associated with the $(1,0)$ $7$-branes can be determined as follows: for each such $7$-brane with $n$ D$5$-branes ending on it we associate a column with $n$ boxes. Starting from highest $n$ to lowest we can combine them to form a Young diagram which is the one associated with the punctures (see figure \ref{Rd0}). It is not difficult to see that if there are no stuck $7$-branes this generates a Young diagram of $D_N$ type on both sides. This also straightforwardly works when there are an even number of stuck $7$-branes on both sides of the $O5$-plane. For an odd number of stuck $7$-branes, besides not giving a $D$ type Young diagram, in this case the number of D$5$-branes on both sides is different. Nevertheless, we can still associate a $4d$ class S theory with this case by moving one of the stuck $7$-branes from the side with more D$5$-branes to the other, doing Hanany-Witten transitions when necessary. Note however that this may change the theory by the generation of free hypermultiplets. We refer the reader to the appendix for details.

This leaves the question of the very-even partitions which should give two different punctures. We find that we can only get one type of puncture. This appears to be related to the question of changing the $\theta$ angle for $USp$ groups engineered by webs in the presence of $O5$-planes (see \cite{Zaf3}). Presumably, if this problem can be resolved then one can also realize the other choice of puncture in the $5d$ web.

\begin{figure}
\center
\includegraphics[width=0.8\textwidth]{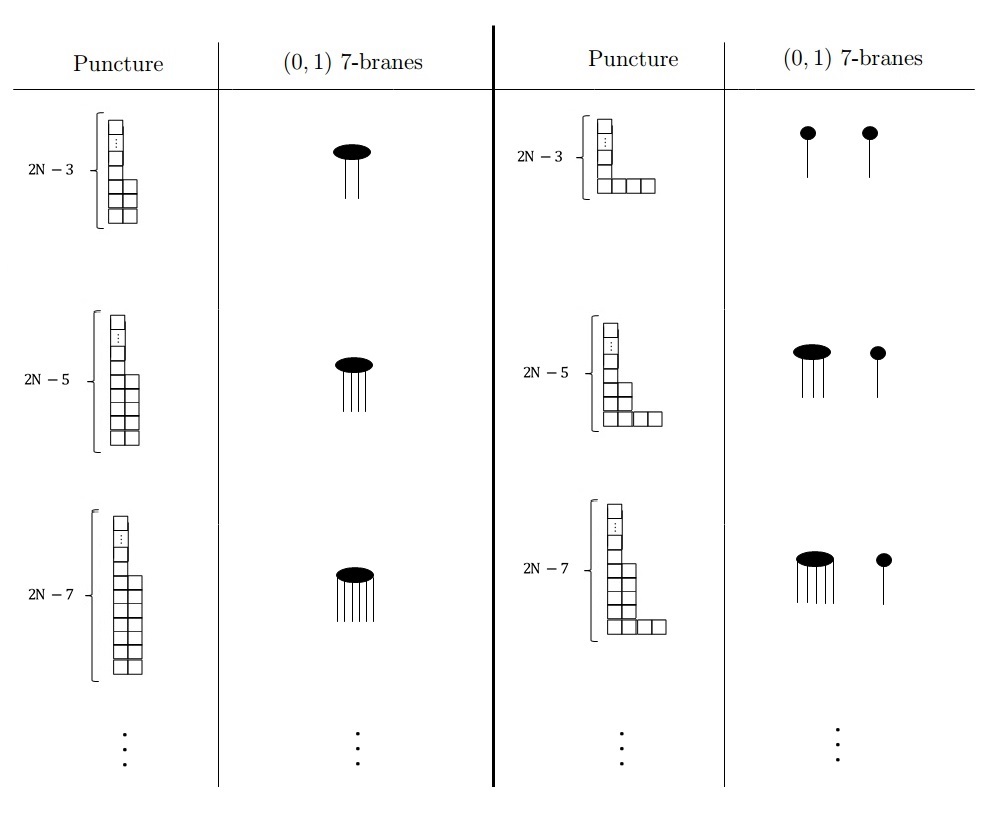} 
\caption{The mapping between the allocation of NS$5$-branes on the $(0,1)$ $7$-branes and the associated puncture. In all cases the $6d$ theory is the $(2,0)$ theory of type $D_{N}$ where $N$ is determined as explained in the text.}
\label{Rd}
\end{figure}

The punctures associated with the $(0,1)$ $7$-branes are shown in figure \ref{Rd}. Basically if there is a single $(0,1)$ $7$-brane with $2n$ NS$5$-branes ending on it then the corresponding puncture is given by a Young diagram with two columns of length $2N-2n-1$ and $2n+1$. Alternatively, if there are two NS$5$-branes, one with a single NS$5$-brane ending on it and one with $2n-1$ NS$5$-branes ending on it, then the Young diagram associated with this puncture has four columns, one of length $2N-2n-1$, another of length $2n-1$ and two of length $1$. 

Next we want to provide evidence for this conjecture. First, note that the global symmetry generated by the $(1,0)$ $7$-branes is in accordance with those given by the punctures. As discussed in the previous section, a group of $k$ D$7$-branes with an even (odd) number of $5$-branes ending on it contributes a $USp(k)$ ($SO(k)$) global symmetry. This indeed agrees with the global symmetry associated with the punctures. 


This usually also works for the $(0,1)$ $7$-branes. First note that one $U(1)$ global symmetry in the web, identified with translations along the $O5$-plane, decouples from the SCFT. So a single $(0,1)$ $7$-brane has no associated global symmetry while two $(0,1)$ $7$-branes have an associated $U(1)$ global symmetry which is enhanced to $SU(2)$ when the number of $5$-branes ending on them is equal. This indeed agrees with the global symmetry expected from the punctures with the exception of the case where the two longest columns in the Young diagram are equal. Then there is an extra $U(1)$ in the puncture, not appearing in the web. 

We can also compare the Higgs branch dimension. As discussed in the previous section, we can break the Higgs branch of the web to $4$ contributions. Likewise we can break the Higgs branch of the $4d$ SCFT to the $4$ contributions appearing in equation (\ref{DoHB}). First there are the two contributions coming from breaking the D$5$-branes on the D$7$-branes on the two sides of the orientifold. The number of possibilities depends on the number of D$5$-branes ending on each $7$-brane and so on the choice of puncture. We identify these with $f_1$ and $f_2$ in (\ref{DoHB}), where there are two for the two sides of the $O5$-plane.


Next we need to show that these two quantities indeed agree, at least for the maximal puncture. Using (\ref{HBfP}) we find that:

\be
f_{Max} = N (N-1)
\ee

Our mapping instructs us to map this to $2N$ $(1,0)$ $7$-branes with one D$5$-brane ending on each one. The number of possible separations is:

\be
2(\sum^{N-1}_{i=1} i) = N (N-1)
\ee
agreeing with the class S result.

 A second contribution comes from pairwise separating the $2N$ D$5$-branes stretching between the two stuck D$7$-branes closest to the NS$5$-branes. The number of possible separations is always $N$, and is identified with the $N$ in equation (\ref{DoHB}). The final contribution comes from breaking the NS$5$-branes across the $(0,1)$ $7$-branes, and pairwise separating them from the $O5$-plane. The number of possibilities depends on the number of NS$5$-branes ending on each $7$-brane and so on the choice of the third puncture. It is thus natural to identify this with $f_3$. Using equation (\ref{HBfP}) we can evaluate $f_3$ for the punctures appearing in figure \ref{Rd}, and we find that these indeed agree to what is expected from the web. For example consider the minimal puncture, which is the left uppermost puncture in figure \ref{Rd}. Using (\ref{HBfP}) we find that $f^{min}_3 =1$. This indeed agrees with what is expected from the web, corresponding to the $1$ dimensional Higgs branch given by pairwise separating the two NS$5$-branes from the $O5$-plane, along the $(0,1)$ $7$-brane. It is straightforward to carry this over also to the other cases. 

So far we gave some rather general arguments, but next we want to put more stringent tests on this conjecture. Consider a $5d$ gauge theory which can be engineered by a web of the previously given form. In some cases this gauge theory may also be constructed by a brane web system without the orientifold. In that case the results of \cite{BB} suggest that compactifying it to $4d$ should lead to a class S theory of type $A$. However, our conjecture implies that compactifying the same theory, in the same limit, should yield a class S theory of type $D$. Consistency now requires that these in fact be the same SCFT which we can test using class S technology. Particularly, we can compute and compare their spectrum and dimensions of Coulomb branch operators, Higgs branch dimension, global symmetry, and central charges of the non-abelian global symmetries. If our conjecture is correct then these must match. 

\begin{figure}
\center
\includegraphics[width=1\textwidth]{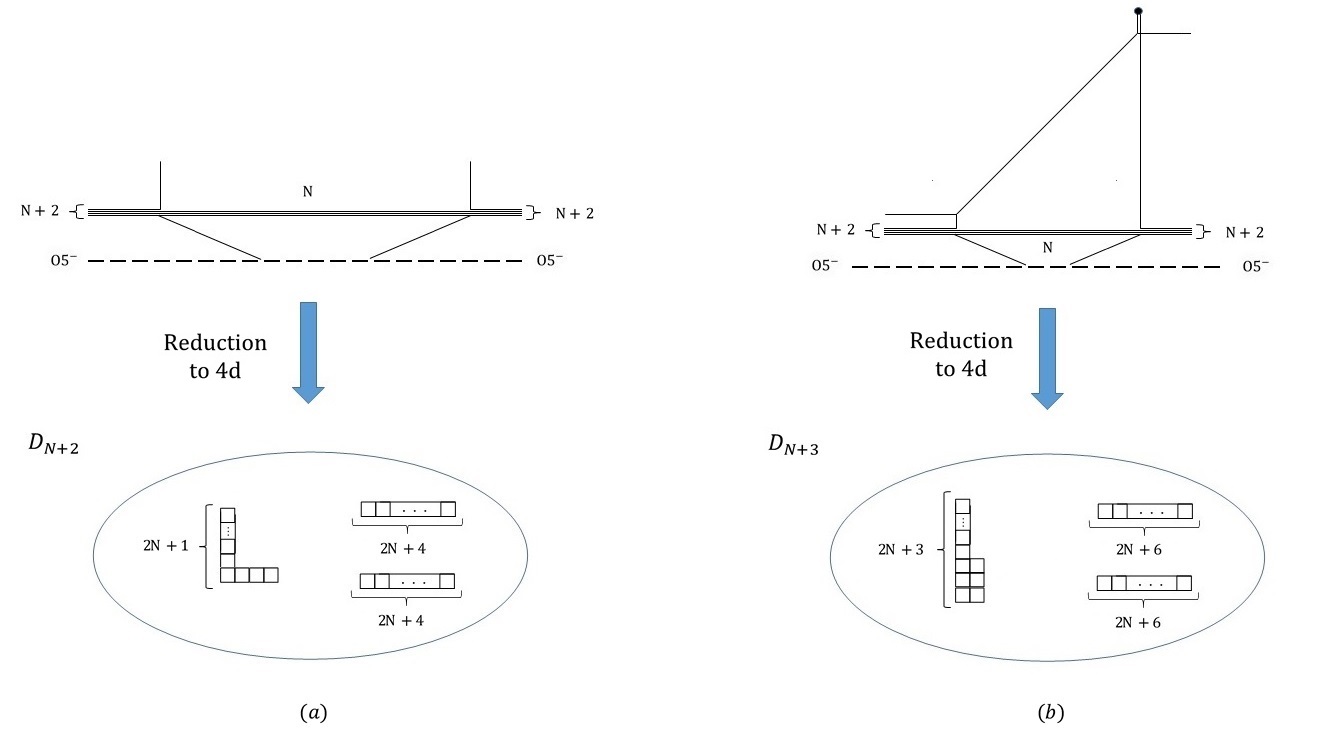} 
\caption{Constructing the $5d$ gauge theories $USp(2N)+(2N+4)F$ (a) and $USp(2N)+(2N+5)F$ (b) using an $O5$-plane. The upper part shows the webs describing the $5d$ SCFT's. These webs are of the form we consider and so we conjecture that their compactification on a circle results in the $D$ type class S theories shown below them.}
\label{Rd11}
\end{figure}

\begin{figure}
\center
\includegraphics[width=1.05\textwidth]{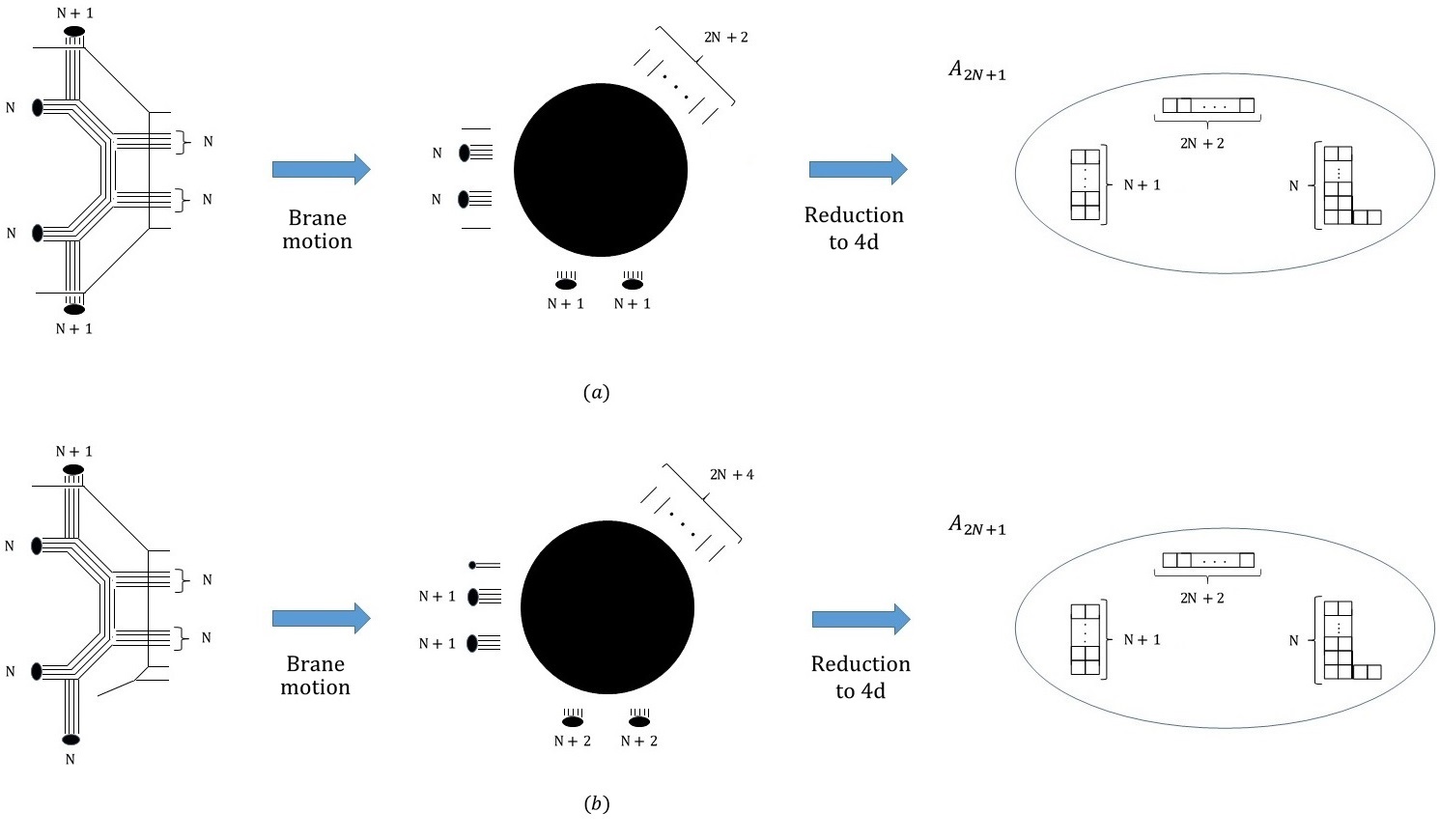} 
\caption{Constructing the $5d$ gauge theories $USp(2N)+(2N+4)F$ (a) and $USp(2N)+(2N+5)F$ (b) using a resolved $O7^-$-plane. The left part shows the webs one gets after resolving the $O7^-$-plane. By performing a series of $7$-brane motions these can be recast into the the form of \cite{BB}. The resulting webs are shown in the middle part of the figure. Upon compactification to $4d$, these are expected to give the type $A$ class S theories shown on the right.}
\label{Rd12}
\end{figure}

There are several types of theories where this can be done. One type are gauge theories of the form $USp(2N)+N_fF$. When $N_f=2N+4, 2N+5$ the webs using an $O5$-plane are indeed of the previously considered form, and so are conjectured to lead to a $4d$ $D$ type class S theory (see figure \ref{Rd11}). These theories can also be engineered using an $O7^-$ plane, which when decomposed leads to an ordinary brane web description\cite{BZ1}. Precisely when $N_f=2N+4, 2N+5$ these webs are of the form of \cite{BB}, and so lead to an $A$ type class S theory (see figure \ref{Rd12}). Note that the identification of the theories in figures \ref{Rd11} (b) and \ref{Rd12} (b) was already done in \cite{Zaf2} while the identification of the theories in figures \ref{Rd11} (a) and \ref{Rd12} (a) is expected from the results of \cite{CDT1,Tachi2}.



\begin{figure}
\center
\includegraphics[width=0.75\textwidth]{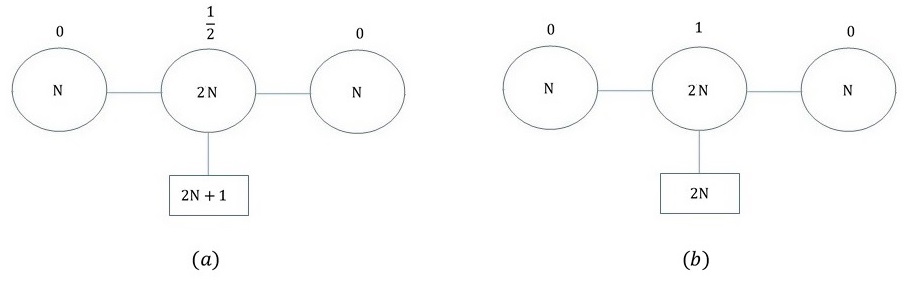} 
\caption{The quiver diagrams for the $5d$ gauge theories whose webs are constructed in figures \ref{Rd22} and \ref{Rd23}. All groups are of type $SU$ with the CS level written above the group.}
\label{Rd21}
\end{figure}

\begin{figure}
\center
\includegraphics[width=1.05\textwidth]{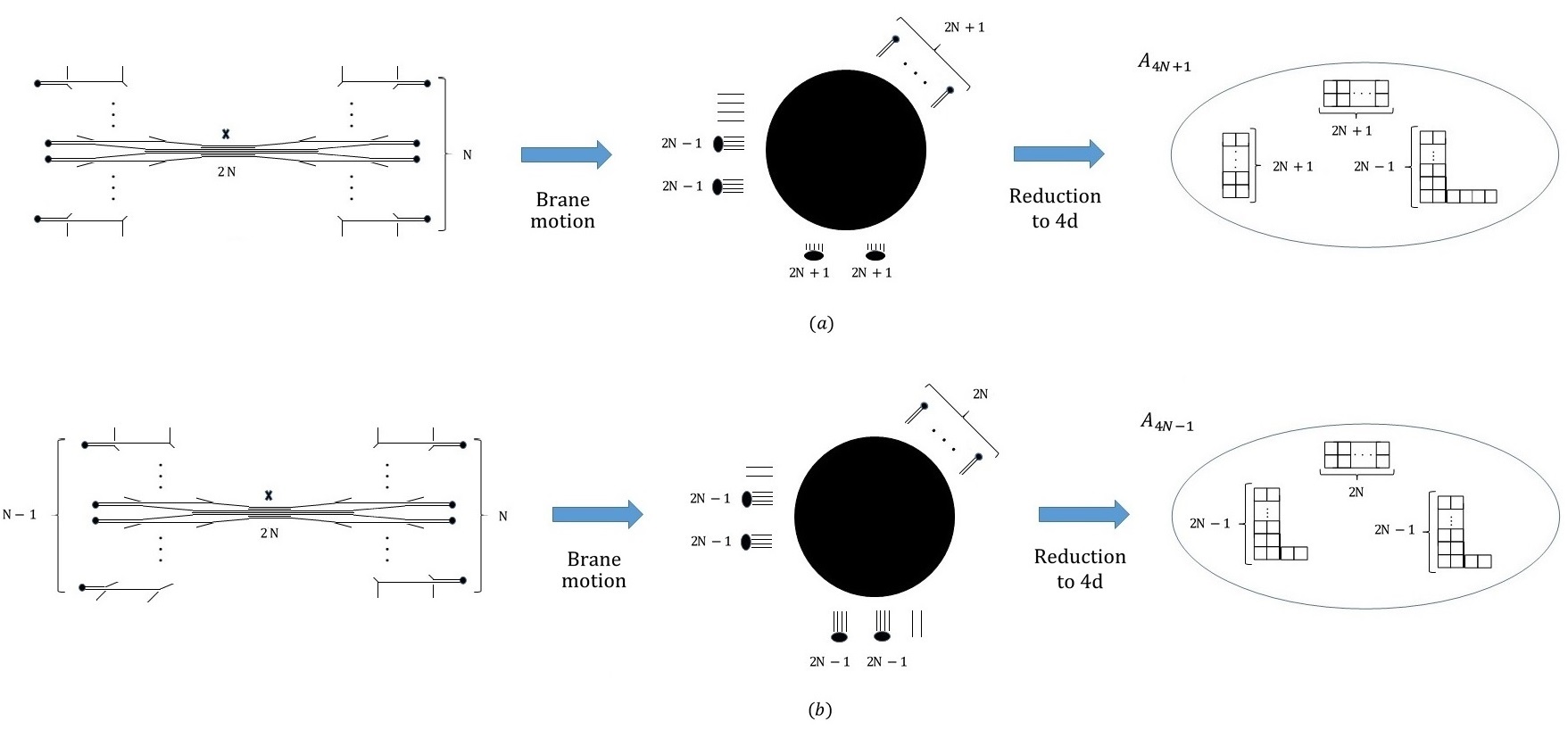} 
\caption{(a) The brane web for the $5d$ gauge theory shown in figure \ref{Rd21} (a). (b) The brane web for the $5d$ gauge theory shown in figure \ref{Rd21} (b). The left part shows the webs in a form where the gauge theory description is manifest. Note that we use black X's for $(1,0)$ $7$-branes. By performing a series of $7$-brane motions, these can be recast into the the form of \cite{BB}. The resulting webs are shown in the middle part of the figure. Upon compactification to $4d$, these are expected to give the type $A$ class S theories shown on the right.}
\label{Rd22}
\end{figure}

\begin{figure}
\center
\includegraphics[width=1.05\textwidth]{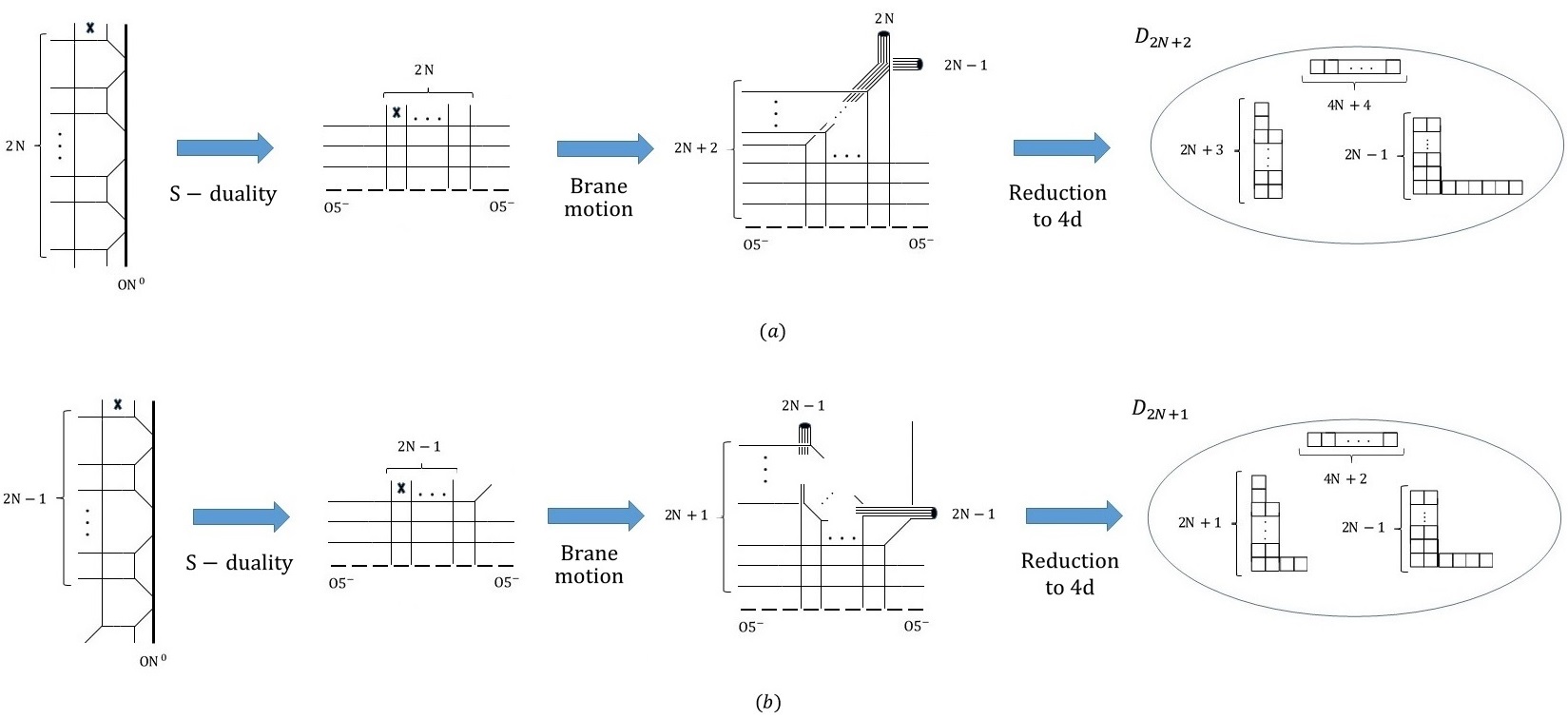} 
\caption{(a) The brane web for the $5d$ gauge theory shown in figure \ref{Rd21} (a), now using an $ON^0$-plane. (b) The brane web for the $5d$ gauge theory shown in figure \ref{Rd21} (b), now using an $ON^0$-plane. The left part shows the webs using the $ON^0$-plane, which makes the $D_3$ shaped quiver nature of the gauge theories manifest. Again we use black X's for $(1,0)$ $7$-branes. Performing S-duality leads a web using an $O5$-plane where for ease of presentation we have suppressed the bending due to the change in $O5$-plane type. By performing a series of $7$-brane motions these can be recast into the form we consider. Therefore we conjecture that their compactification on a circle results in the $D$ type class S theories shown on the right.}
\label{Rd23}
\end{figure}

 As mentioned in the previous section, the S-dual of the webs considered here generally leads to $D$ shaped quivers of $SU$ groups. In the case of $D_3$ the $D$ shape degenerates to a linear quiver which can also be engineered using ordinary brane webs. This provides another case where we can compare two different constructions of the same gauge theory, where one is expected to give an $A$ type class S theory, while the other a $D$ type. Figures \ref{Rd21}, \ref{Rd22} and \ref{Rd23} show two examples of theories in this class. Figure \ref{Rd21} shows the quiver diagram for the $5d$ gauge theories we consider. Figure \ref{Rd22} then shows the realization of these gauge theories using brane webs without orientifolds, which is possible since these are linear quivers of $SU$ groups. Finally figure \ref{Rd23} shows the realization of these gauge theories using brane webs in the presence of an $ON^0$-plan, which is possible as these are $D_3$ shaped quivers. This can be mapped to a configuration with an $O5^-$-plane using S-duality.


we can test these identifications by comparing the spectrum and dimensions of Coulomb branch operators, Higgs branch dimension, global symmetry, and central charges of the non-abelian global symmetries of the two SCFT's\footnote{We can also compare the $a$ and $c$ conformal anomalies of the SCFT's. However, for class S theories these are completely determined in terms of the spectrum and dimensions of Coulomb branch operators and the dimension of the Higgs branch\cite{CD}, and so do not provide an independent check.}. For the $A$ type class S theory in figure \ref{Rd22} (a) we find using class S technology that it has a Higgs branch dimension of $d_H = 4N^2 + 10N + 15$ and global symmetry $SO(4N+6)_{8N+4}\times SU(4)_{12}$ where the subscript denote the central charge. We also find the spectrum of Coulomb branch operators to be:

\bea
d_2 & = & d_3 =0, d_4 = 1 \nonumber \\
d_i & = &  
\left \{
  \begin{tabular}{cc}
  2 & i \text{even}\\
  1 & i \text{odd}
  \end{tabular}
\right \} 
\text{ for } 4<i\leq 2N+1 \nonumber \\
d_i & = &  
\left \{
  \begin{tabular}{cc}
  1 & i \text{even}\\
  0 & i \text{odd}
  \end{tabular}
\right \} 
\text{ for } 2N+1<i\leq 4N+2 \nonumber
\eea 
where $d_i$ stands for the number of such operators with dimension $i$, and we assumed $N>1$ (for $N=1$ the matching reduces to the $N=1$ case of figures \ref{Rd11} (b) and \ref{Rd12} (b)). We can now compare these values with the ones for the $D$ type class S theory in figure \ref{Rd23} (a). Using class S technology we carried out this calculation finding complete agreement.  

For the $A$ type class S theory in figure \ref{Rd22} (b) we find using class S technology that it has a Higgs branch dimension of $d_H = 4N^2 + 4N + 7$ and global symmetry $SO(4N+4)_{8N}\times SU(2)^2_{8}\times U(1)$. We also find the spectrum of Coulomb branch operators to be:

\bea
d_2 & = & 0 \nonumber \\
d_i & = &  
\left \{
  \begin{tabular}{cc}
  2 & i \text{even}\\
  1 & i \text{odd}
  \end{tabular}
\right \} 
\text{ for } 2<i\leq 2N \nonumber \\
d_i & = &  
\left \{
  \begin{tabular}{cc}
  1 & i \text{even}\\
  0 & i \text{odd}
  \end{tabular}
\right \} 
\text{ for } 2N<i\leq 4N \nonumber
\eea 

We can now compare these values with the ones for the $D$ type class S theory in figure \ref{Rd23} (b). Using class S technology we carried out this calculation finding complete agreement.

\begin{figure}
\center
\includegraphics[width=1.1\textwidth]{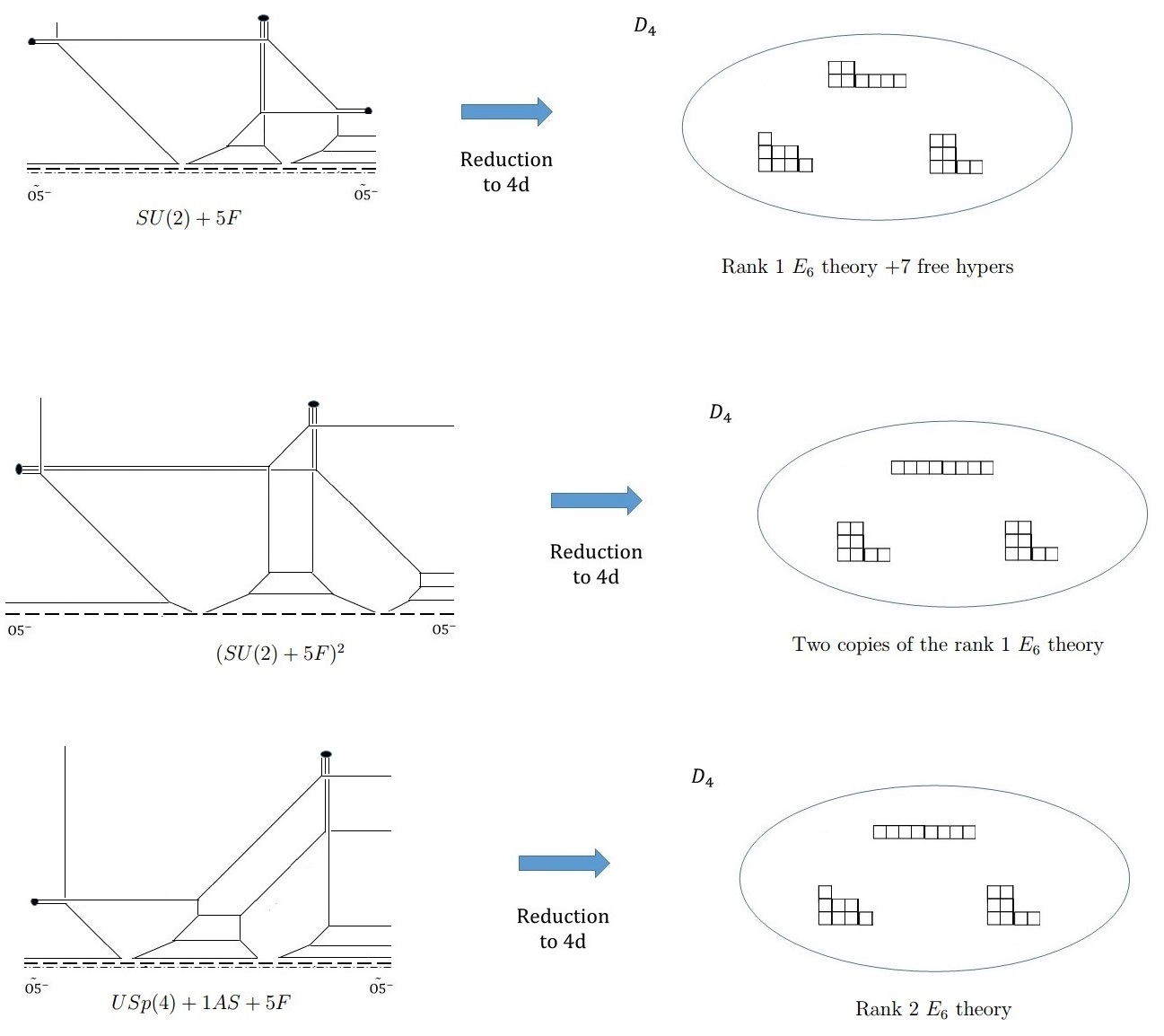} 
\caption{Engineering $E_6$ type theories using an $O5$-plane. Shown are $3$ examples including their associated class S theories being a compactification of the $D_4$ theory where we have used that $SO(3)=SU(2)$, $SO(4)=SU(2)^2$ and $SO(5)=USp(4)$. Up to the free hypers in the first case, these match what is expected from the gauge theory description, see \cite{CD1}.}
\label{Rd3}
\end{figure}

\begin{figure}
\center
\includegraphics[width=1.1\textwidth]{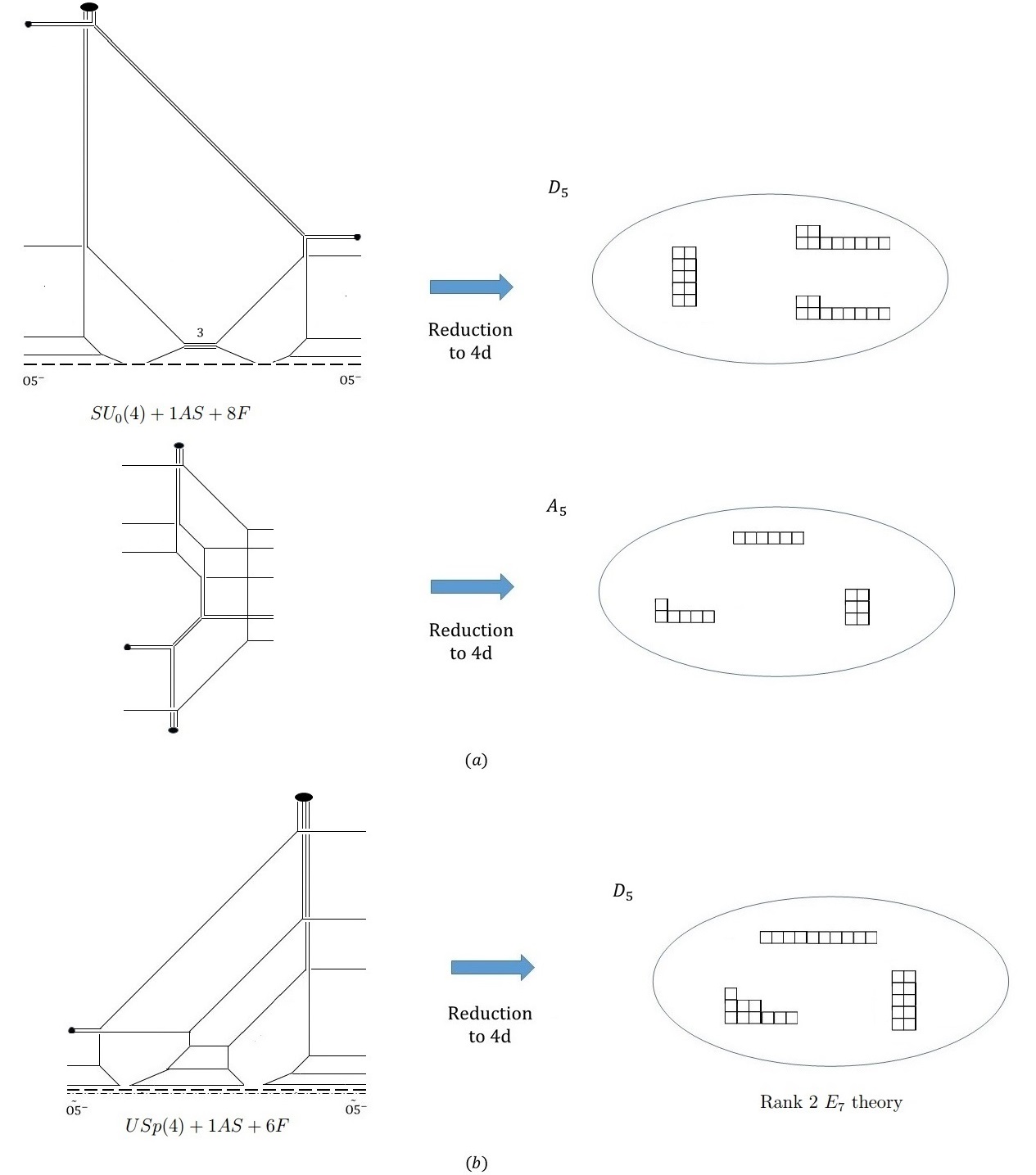} 
\caption{Two examples for the $D_5$ theory. (a) Shows the case of $SU_0(4)+1AS+6F$, with the associated $D_5$ theory and the construction using ordinary brane webs leading to an $A_5$ theory. (b) Shows another example now using the rank $2$ $E_7$ theory.}
\label{Rd4}
\end{figure}

We can also consider $SO(N)$ gauge theories with spinor matter for $N\leq 6$. In that range the theory can also be engineered using brane webs without the orientifold. Selected examples are shown in figure \ref{Rd3}, for cases with a $D_4$ class S theory, and figure \ref{Rd4} for cases with a $D_5$ class S theory. All the class S theories of type $D_4$ were analyzed in \cite{CD1}, and their results indeed agree with our expectation from the $5d$ SCFT. 

It is straightforward to also carry this analysis for the theories in figure \ref{Rd4}. For the $D$ type class S theory in figure \ref{Rd4} (a) we find it has an $SU(4)_{10} \times SU(8)_{12}$ global symmetry, a $33$ dimensional Higgs branch and Coulomb branch operators of dimensions $4,5,6$. This indeed matches the properties of the $A$ type class S theory in figure \ref{Rd4} (a) evaluated using class S technology. We can do the same for the theory in figure \ref{Rd4} (b) now finding an $SU(2)_{9} \times (E_7)_{16}$ global symmetry, a $35$ dimensional Higgs branch and Coulomb branch operators of dimensions $4,8$. This indeed matches the properties of the rank $2$ $E_7$ theory as can be computed using class S technology from the construction of this theory with a free hyper given in \cite{GR}.  

We do note that sometimes the class S theory contains free hypers in excess of those expected from $7$-brane motions (as explained in the appendix). This seems common for $SO(3)$ and $SO(4)$, and an example for this is given by the first case in figure \ref{Rd3}. Nevertheless, the Higgs branch of the class S theory with the free hypers agrees with that of the web. Thus, the natural interpretation is that the SCFT described by the web in these cases also contains free hypers. As mentioned in \cite{Zaf3}, the spinors are thought to come from instantons of an $SU(2)$ gauge theory broken by a motion on the Higgs branch. Thus, it is tempting to identify these free hypers as directions on that Higgs branch that somehow remain in the low-energy theory.    


\begin{figure}
\center
\includegraphics[width=0.9\textwidth]{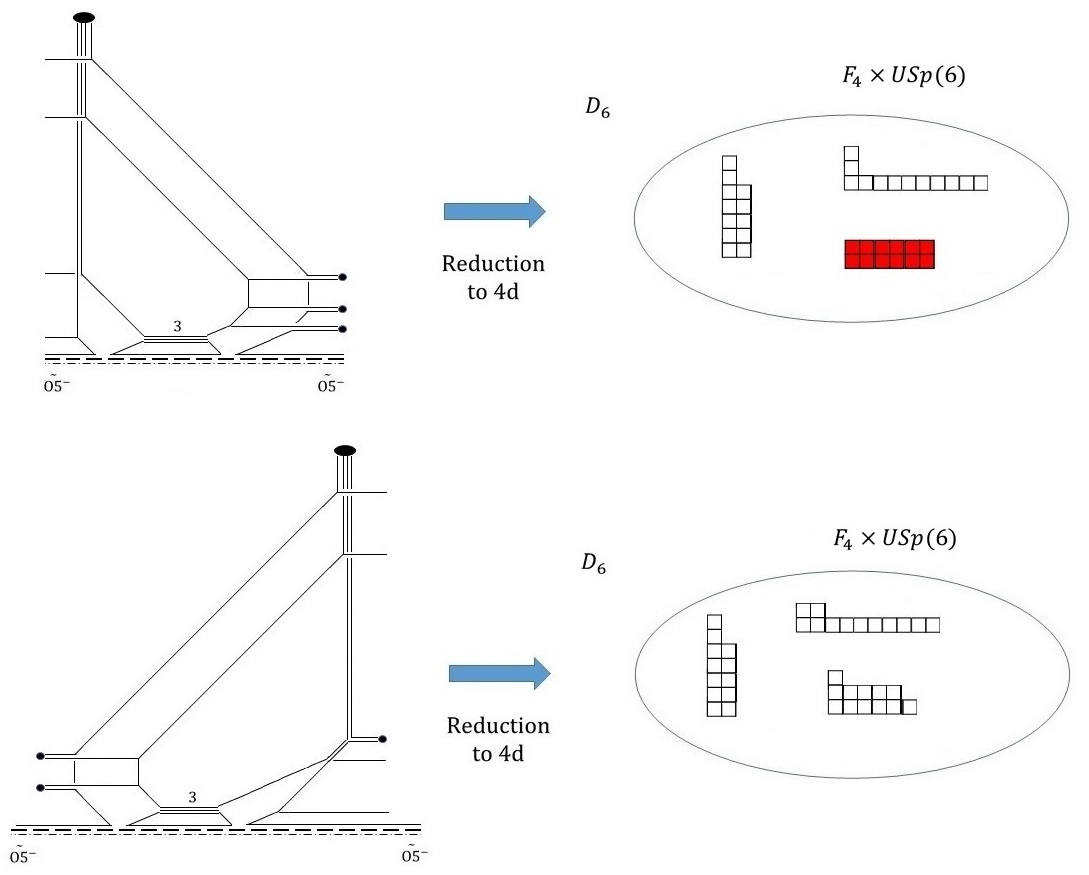} 
\caption{Two webs and the related class S theories for the $5d$ gauge theory $SO(7)+3V+3S$. The webs are shown on the left, and upon moving the rightmost stuck D$7$-brane to the leftmost side of the web, they can be cast in the form we consider. Note that this $7$-brane motion results in the creation of free hypers. We conjecture that compactification on a circle to $4d$ results in the $D$ type class S theories shown on the right. Above each class S theory we have written its global symmetry evaluated using the superconformal index. These agree with the expected global symmetry of the $5d$ SCFT determined from the $1$ instanton analysis of \cite{Zaf1}.}
\label{Rd51}
\end{figure}

\begin{figure}
\center
\includegraphics[width=0.9\textwidth]{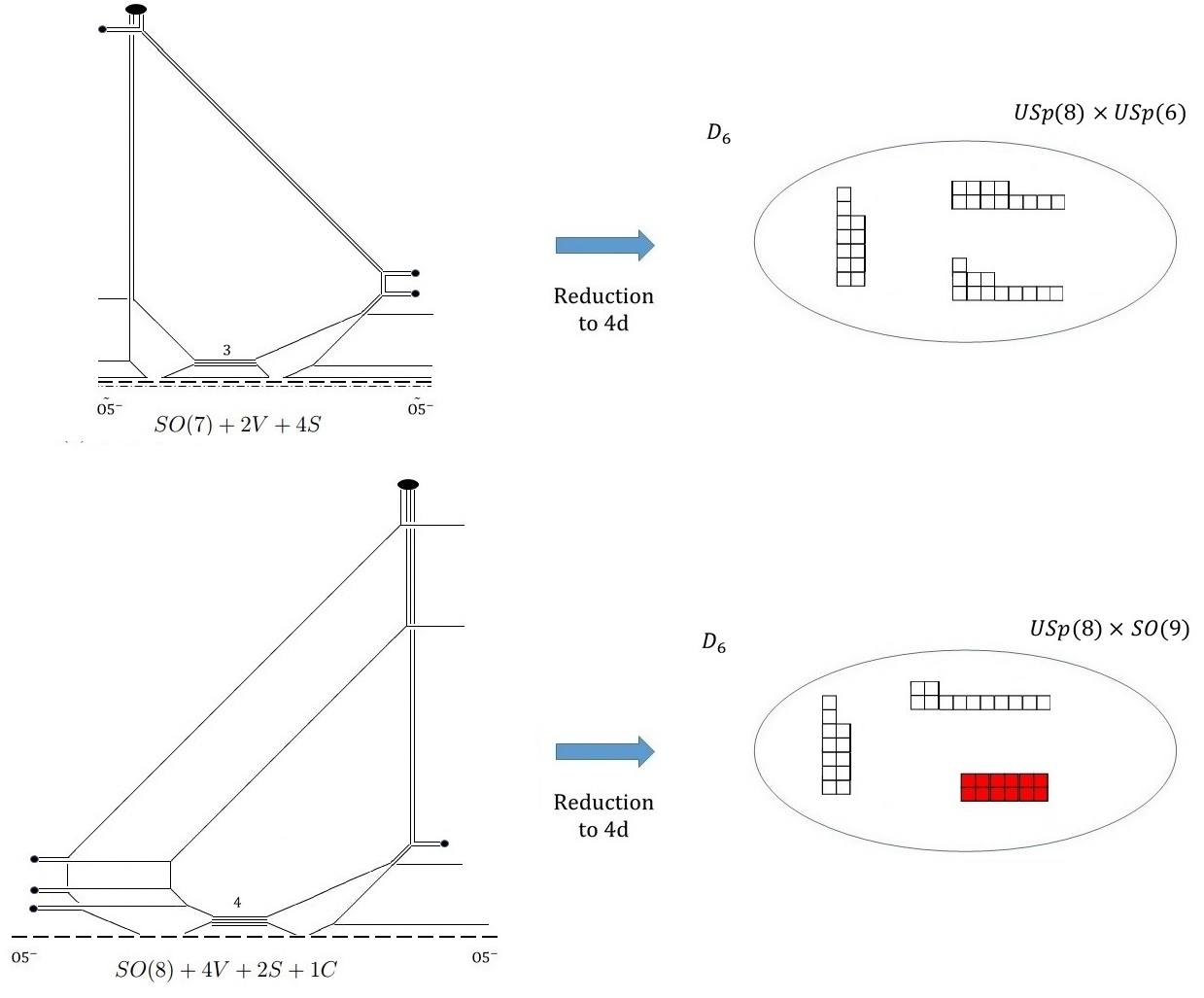} 
\caption{Two webs and the related class S theories for the $5d$ gauge theories $SO(7)+2V+4S$ (upper part) and $SO(8)+4V+2S+1C$ (lower part). The webs are shown on the left, where the lower one is already in the form we consider. The upper web can also be cast in this form by moving the rightmost stuck D$7$-brane to the leftmost side of the web, though this leads to the creation of free hypers. We conjecture that compactification on a circle to $4d$ results in the $D$ type class S theories shown on the right. Above each class S theory we have written its global symmetry evaluated using the superconformal index. These agree with the expected global symmetry of the $5d$ SCFT determined from the $1$ instanton analysis of \cite{Zaf1}.}
\label{Rd52}
\end{figure}

A further consistency check is to match the global symmetry expected from the $5d$ description to the one of the $4d$ SCFT calculated using class S technology. As an example, consider an $SO(N)$ gauge theory with spinor matter. In \cite{Zaf1} an analysis of the $1$ instanton operators was done for these theories, which in turn suggests that enhancement of symmetries should occur in some cases. The class S theories which we conjecture result from compactifying these theories do not show that symmetry, and in many cases not even the classically visible symmetry. Consistency now requires that the $4d$ global symmetry is actually enhanced to a larger global symmetry than is visible from the punctures, which we can test in turn using class S technology. Several examples are shown in figures \ref{Rd51} and \ref{Rd52}, where in all cases complete agreement with the results of \cite{Zaf1} is found, and further that the Coulomb branch dimension of the class S theory agrees with that expected from the web. Also class S technology calculations are consistent with the two theories in figure \ref{Rd51} being the same theory up to a differing number of free hypers.

\subsection{SCFT's with marginal deformations}      

Next we want to discuss theories related to the compactification of the $D$ type $(2,0)$ theory on a sphere with more than three punctures. Unlike the previous case the reduction now involves several scaling limits where in addition to taking the $R_5\rightarrow 0$ limit one also takes the limit $m_i\rightarrow \infty$, for several mass parameters $m_i$, while keeping $m_i R_5$ fixed. These then become the marginal deformations that exist in compactification of a $(2,0)$ theory on a sphere with more than three punctures. These mass parameters can be chosen as $m_i = \frac{1}{g^2_{5d}}$ corresponding to the couplings of gauge groups. These then become marginal gauge group couplings in $4d$ which are indeed related to their $5d$ counterparts by $\frac{1}{g^2_{4d}} \sim \frac{R_5}{g^2_{5d}}$.

The actual correspondence is a straightforward generalization of the previous case. The theories in this class are of the same form as the previous ones except that the number of $(0,1)$ $7$-branes is unconstrained save for the demand that they can be partitioned into groups each one in the form of figure \ref{Rd}. We can then do the reduction in the limit where each group of $(0,1)$ $7$-branes is infinitely separated from the other. This corresponds to a weakly coupled limit for the gauge theory living on the D$5$-branes stretched between neighboring groups of $(0,1)$ $7$-branes. 

It is thus straightforward to conjecture that the resulting theory is given by a collection of isolated SCFT's connected by the gauge groups whose weak coupling limit was taken. This in turn is equal to the compactification of a $D$ type $(2,0)$ theory on a Riemann sphere with punctures where each group of $(0,1)$ $7$-branes and the $(1,0)$ $7$-branes on the two sides of the $O5$-plane gives a puncture of the type explained in the previous subsection.

\begin{figure}
\center
\includegraphics[width=1.0\textwidth]{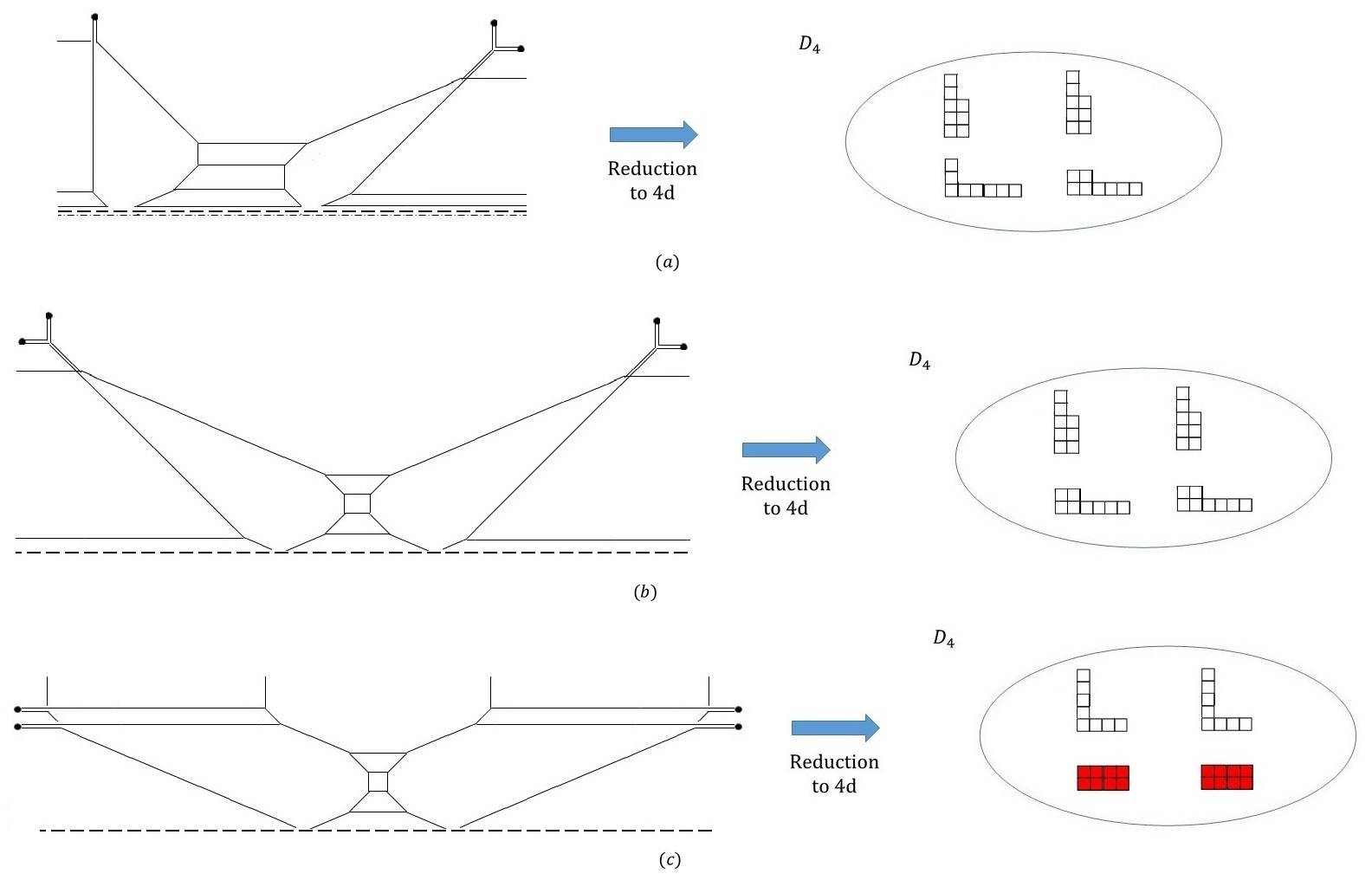} 
\caption{The web description for $SO(7)+4S+1V$ (in (a)), $SO(8)+2V+2S+2C$ (in (b)) and $SO(8)+4V+2S$ (in (c)). These are in a form that we conjecture that when reduced to $4d$, in the infinite weakly coupled limit, lead to a compactfication of the $D_4$ theory on the shown four punctured sphere. Indeed the analogue $4d$ gauge theories are conformal, and as was found in \cite{CD1} have the shown representation as a compactification of the $D_4$ $(2,0)$ theory. Also note that the $4d$ theory in (a) also contains a free hyper that from the $5d$ point of view is generated when moving the right stuck $7$-brane to the left side.}
\label{Rd6}
\end{figure}

As an example we consider $SO$ gauge theories with spinors and vector matter, for example the $SO(8)$ and $SO(7)$ theories shown in figure \ref{Rd6}. With the chosen matter content, the webs are of the form expected to lead to a four punctured sphere when reduced to $4d$. Furthermore it is clear from the web that the reduction is done in the limit where the $SO$ group becomes weakly coupled, and we expect it to descend to the corresponding $SO$ group with marginal coupling. In fact the matter content is exactly the one required for the analogous $4d$ theory to be conformal. So we expect the resulting theory to have a completely perturbative description given by an $SO$ gauge group with vector and spinor matter. 

In fact such $4d$ theories were constructed, using compactification of $D$ type $(2,0)$ theories, in \cite{CD1}, and we can compare the class S construction they give with the one generated from the web by our prescription, finding complete agreement (see figure \ref{Rd6}). This naturally leads to one frame of the many possible frames obtained by taking different pair of pants decompositions of the Riemann surface. One may ask whether the other are also manifest in the $5d$ description. Some of them, the ones corresponding to exchanging the punctures associated with a group of $(0,1)$ $7$-branes, are manifest as we can exchange them also in the web. This leads to $5d$ lifts of Gaiotto dualities, similar to the cases studied in \cite{BZ} for the $A$ type theories.



In the $4d$ theory we can exchange any of the punctures while in the $5d$ theories there does not appear to be a way to exchange the punctures associated with the $(0,1)$ $7$-branes with the ones associated with the $(1,0)$ $7$-branes. This is already apparent from the mapping, since we can essentially get any puncture from the $(1,0)$ $7$-branes, but only a subset from the $(0,1)$ $7$-branes. It would be interesting to find a generalization of these mappings allowing a description of all types of punctures also with the $(0,1)$ $7$-branes.  

Finally we wish to discuss the issue of very-even partitions. To do this we draw your attention to figure \ref{Rd6} (c), in which there is an example of a class S theory with two very-even partitions. In these cases there are two possible theories depending on whether the two very-even partitions are of the same or differing colors. In both cases the theory is an $SO(8)$ gauge theory with four hypermultiplets in the vector representation and two in the spinor. The spinors have the same chirality if the colors are the same, but different otherwise. So one can see that the choice of very-even partition is related to the chirality of the spinors. It is currently unclear how one can change the chirality of the spinors in the web, and so how the other choice of very-even partition can be realized. See \cite{Zaf3} for a discussion of this point. 

\section{Conclusions}

In this article we have proposed a connection between a class of $5d$ SCFT's engineered by brane webs in the presence of $O5$-planes and $4d$ class S theories of type $D$. It will be interesting to further explore this, and see if further evidence for this proposal can be found. 

This has several interesting implications for the class S theories involved. First, the $5d$ theory may have one or more gauge theory descriptions. This implies that the corresponding class S theory has mass deformations leading to the analogous $4d$ gauge theory. We have also seen that this proposal motivates identifications of apparently distinct class S theories. Using the work of \cite{BTX}, this then also implies an identification of the dual mirror theories, which may lead to interesting $3d$ dualities.  

Also there are several other intriguing questions that warrant further exploration. We have seen that there are class S theories of type $D$ for which we do not know the $5d$ lift, notably the $D_N$ theory. It is interesting if there is some generalization of these constructions that may allow the realization of the $5d$ lifts of these theories as well. On the other hand, there are other brane webs system with an $O5$-plane that are not of the desired form. It will be interesting to also study their $4d$ limit.

\subsection*{Acknowledgments}

I would like to thank Oren Bergman and Shlomo S. Razamat for useful comments and discussions. G.Z. is supported in part by the Israel Science Foundation under grant no. 352/13, and by the German-Israeli Foundation for Scientific Research and Development under grant no. 1156-124.7/2011.

\appendix

\section{Brane motions and free hypermultiplets}

\begin{figure}
\center
\includegraphics[width=0.8\textwidth]{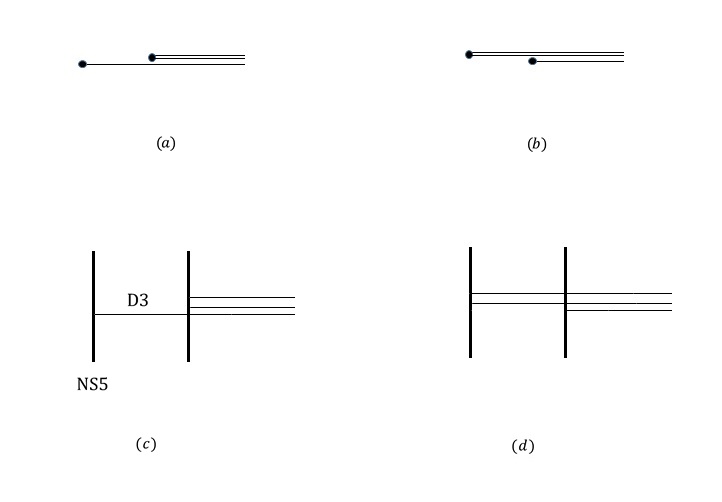} 
\caption{(a) A system of two D$7$-branes with D$5$-branes ending on them. (b) The configuration after moving the two D$7$-branes past each other. (c) The brane configuration one gets after performing two T-dualities in the directions parallel to both the D$7$-branes and the D$5$-branes, followed by an S-duality, on the system of (a). The vertical wide black lines are NS$5$-branes and the thin horiozontal lines are D$3$-branes. (d) The resulting brane configuration after the same transformations are done on the system in (b).}
\label{App1}
\end{figure}

In this appendix we wish to discuss the creation of free hypermultiplets as a consequence of brane motions. Consider the two systems shown in figures \ref{App1} (a) and (b), related by a $7$-brane motion. The system in (a) has a one dimensional Higgs branch, corresponding to breaking the extended D$5$-brane on the $7$-brane. However, in (b) there is now a two dimensional Higgs branch so it appears that this motion has affected the theory. Furthermore, as the Coulomb branch moduli hasn't changed, the most logical explanation is that a single free hyper was created in this motion.

\begin{figure}
\center
\includegraphics[width=1\textwidth]{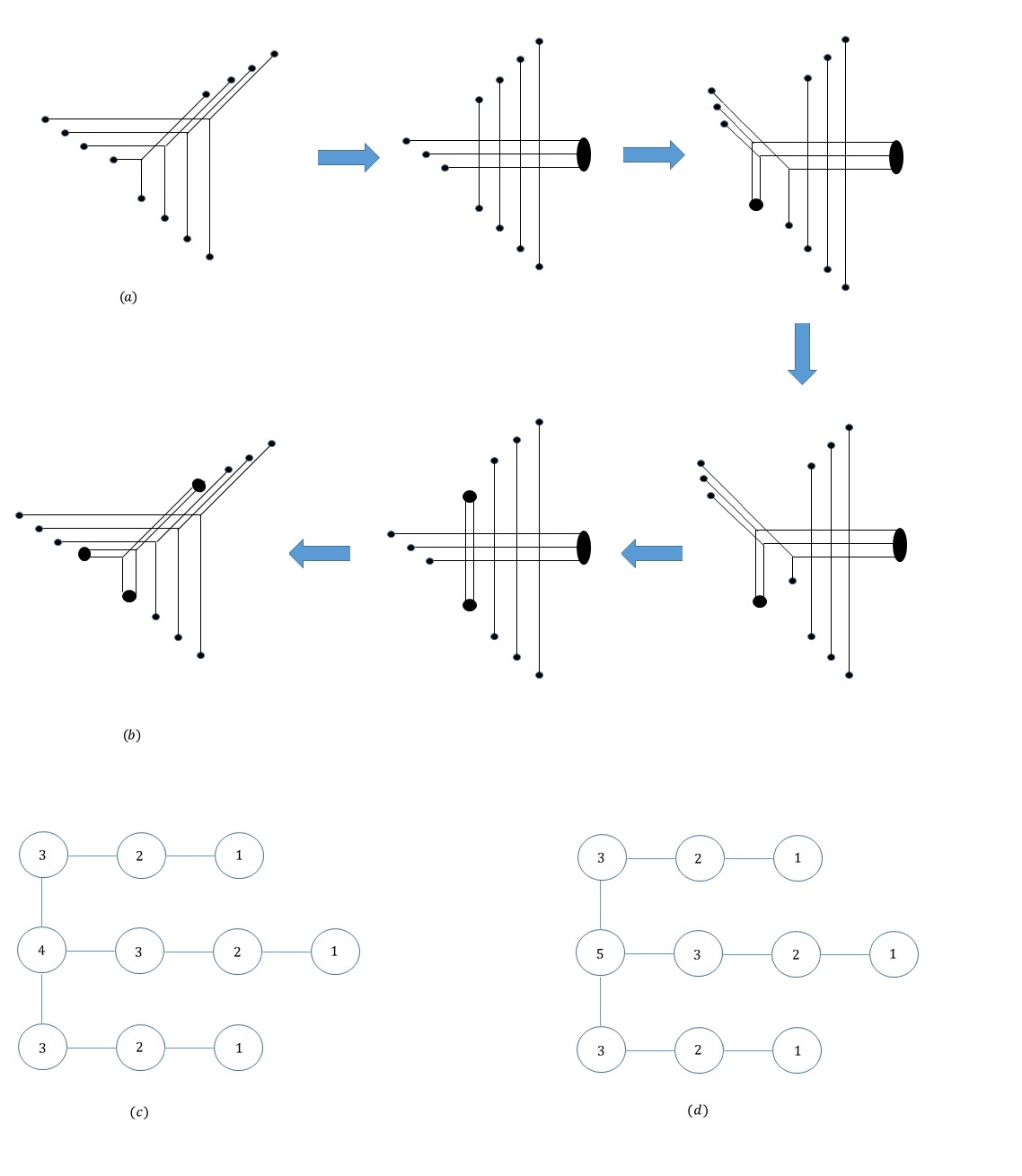} 
\caption{(a) The brane web for the $5d$ $T_4$ theory. After several $7$-branes manipulations we arrive at the configuration in (b). Note that these include a transition of the type shown in figure \ref{App1}. (c) and (d) are the mirror duals of the $3d$ analogue class S theory for the theories in (a) and (b) respectively.}
\label{App2}
\end{figure}

Next, we wish to show that this leads to known relations in other dimensions. Consider taking the systems in figure \ref{App1} (a)+(b), performing two T-dualities and an S-duality, so that the $7$-branes become NS$5$-branes and the D$5$-branes become D$3$-branes. This leads to the brane configuration of figures \ref{App1} (c) and (d) which are again related by brane motion, now of the NS$5$-branes. These motions and the dualities they imply for the theories on the D$3$-branes, where considered in \cite{KYW} who found that indeed the theory in (d) is the one in (c) with the addition of a free twisted hypermultiplet.

As an additional example consider the theory shown in figure \ref{App2} (a). This describes the $5d$ $T_4$ theory. Figure \ref{App2} (b) shows the system after several brane motions which also involves one motion of the type considered before. Thus, we conclude that the theory shown in figure \ref{App2} (b) is the $5d$ $T_4$ theory with a single free hyper. This theory is also in the form of \cite{BB} so it leads to $4d$ class S theory. This theory is indeed known to be $T_4$ with a single free hyper\cite{CD}.

Another interesting test is given by employing the relation of \cite{BTX} between the $4d$ class S theory and the $3d$ mirror dual. These are shown in figure \ref{App2} (c) and (d) for the two theories involved. The preceding imply that these two are identical up to a free twisted hyper, and indeed this is just an application of the previous duality of \cite{KYW} to the central node.    

\begin{figure}
\center
\includegraphics[width=0.8\textwidth]{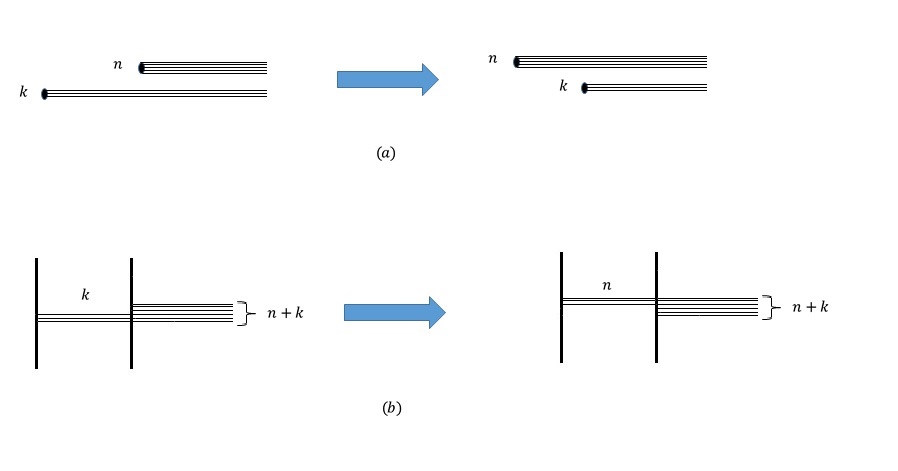} 
\caption{(a) The motion in the general case. (b) The two systems after two T-dualities in the directions parallel to both the D$7$-branes and the D$5$-branes, and an S-duality.}
\label{App3}
\end{figure}

We can ask what happens in the more general case shown in figure \ref{App3} (a). Counting the Higgs branch moduli we find that it changes by $|n-k|$ so the natural expectation is that $|n-k|$ free hypers are generated. We can again consider the related $3d$ system shown in figure \ref{App3} (b). If we are correct then these must be dual up to $|n-k|$ free twisted hypers. This has indeed been argued in \cite{Yak} from $3d$ reasoning. 

\begin{figure}
\center
\includegraphics[width=1.1\textwidth]{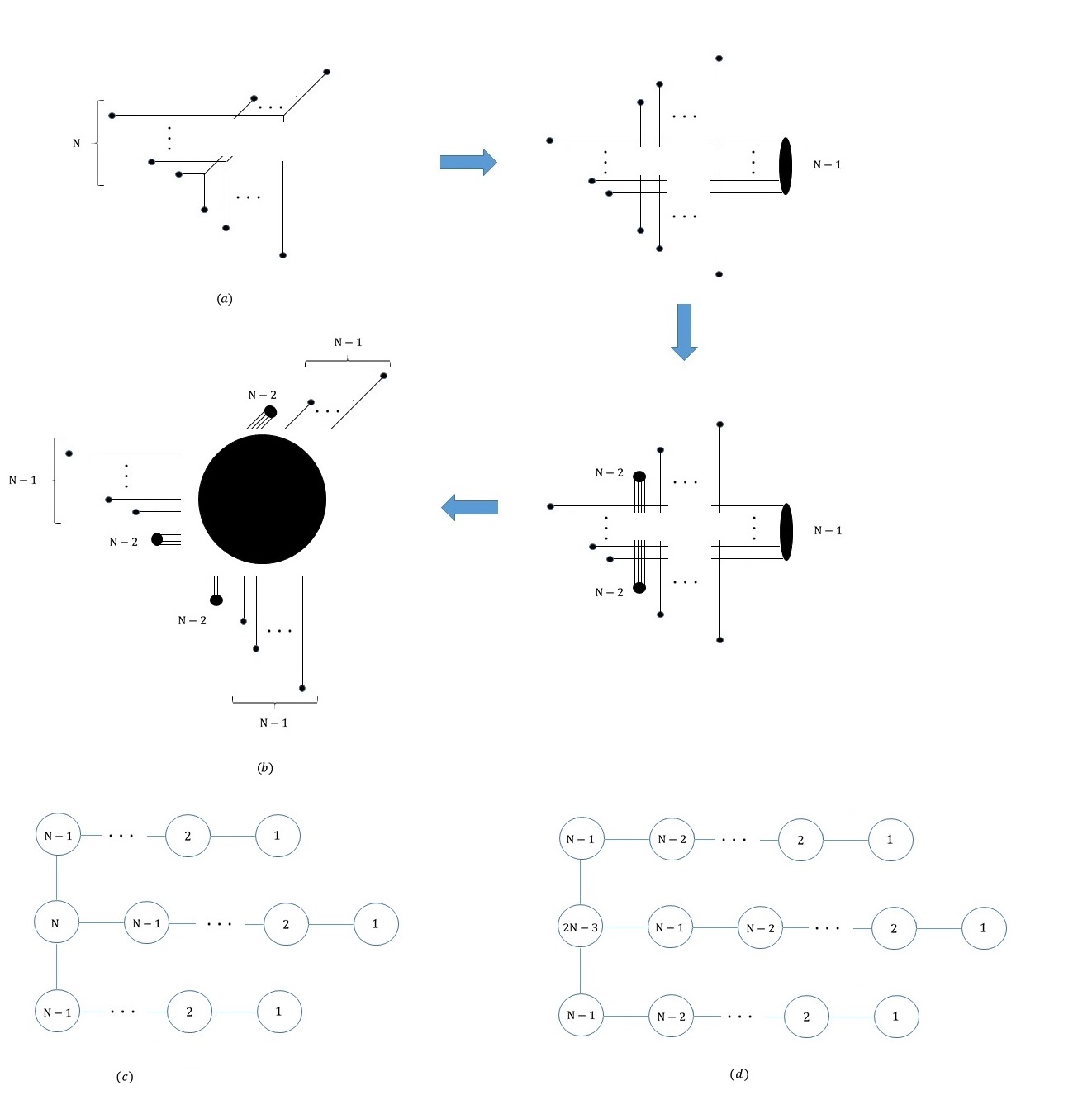} 
\caption{(a) The brane web for the $5d$ $T_N$ theory. Repeating the similar step as done in figure \ref{App2} leads to the web in (b). (c) and (d) are the mirror duals of the $3d$ analogue class S theories for the theories in (a) and (b) respectively.}
\label{App4}
\end{figure}

We can also consider the analogue application for class S theories. Figure \ref{App4} shows similar motions as the ones done for the $T_4$ theory now done for $T_N$. This leads to identifying the two resulting class S theories up to $N-3$ free hypermultiplets. For $N>4$ these theories are bad in the sense of \cite{GR} so index analysis is unknown for them. Also application of class S technology leads to a negative number of Coulomb branch operators and so is not applicable in this case. Yet, the preceding analysis suggests that these theories exist and are given by the $T_N$ theory with free hypermultiplets. One supporting evidence is to consider the $3d$ mirror duals shown in figure \ref{App4}, which are indeed related by the duality of \cite{Yak} on the central node. 

\begin{figure}
\center
\includegraphics[width=0.8\textwidth]{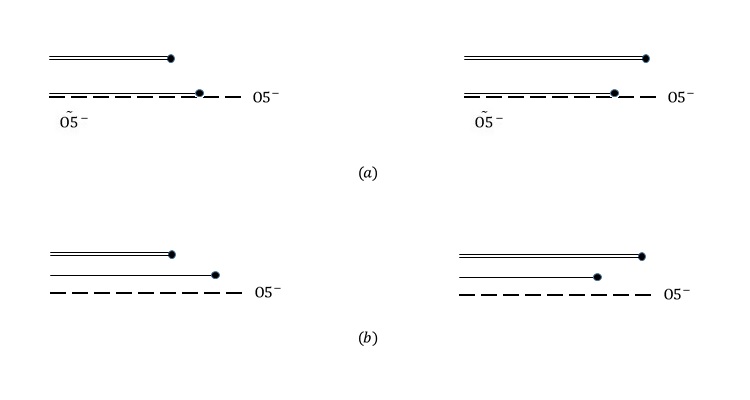} 
\caption{The same steps as in figure \ref{App1}, now in the presence of an $O5^-$ plane. (a) Shows the case where one of the $7$-branes is stuck while (b) shows the case where both are full.}
\label{App5}
\end{figure}

Finally we wish to discuss the implications of these on the systems considered in this article. This phenomenon still occurs with little changes due to the presence of the orientifold plane. For example consider the systems shown in figure \ref{App5} (a) and (b). Counting the Higgs branch dimensions we see that in the case of (a) we get one free hyper while in the case of (b) we get two. It is straightforward to generalize this to other cases.

We next ask whether we can test this also in this case against known phenomena in $4d$ or $3d$. The discussion in section 3 implies that this should lead to similar relationships between different class S theories of type $D$, but it appears at least one of them is bad (again in the sense of \cite{GR}) making explicit comparison difficult. One interesting implication of this is a generalization of the dualities in \cite{Yak} also to $SO$ or $USp$ gauge theories. To our knowledge these have not been previously discussed from purely $3d$ reasoning. It will be interesting to further explore this.


\begin{thebibliography}{40}

\bibitem{SEI}
  N.~Seiberg, 
   Phys.\ Lett.\  B388:753-760 (1996)
  [arXiv:9608111 [hep-th]].

\bibitem{Gai} 
  D.~Gaiotto,
  JHEP {\bf 1208}, 034 (2012)
  [arXiv:0904.2715 [hep-th]].

\bibitem{MN} 
  J.~A.~Minahan and D.~Nemeschansky,
  Nucl.\ Phys.\ B {\bf 482}, 142 (1996)
  [hep-th/9608047], 
    Nucl.\ Phys.\ B {\bf 489}, 24 (1997)
  [hep-th/9610076].

\bibitem{BB}
  F.~Benini, S.~Benvenuti, and Y.~Tachikawa,
  JHEP {\bf 0909}, 052 (2009)
  [arXiv:0906.0359 [hep-th]].

\bibitem{HA}
  O.~Aharony, A.~Hanany,
   Nucl.\ Phys.\  B504:239-271 (1997)
  [arXiv:9704170 [hep-th]].
	
\bibitem{AHK}
  O.~Aharony, A.~Hanany, and B.~Kol,
   JHEP {\bf 9801}, 002 (1998)
  [arXiv:9710116 [hep-th]].


  





\bibitem{BMPTY}
  L.~Bao, V.~Mitev, E.~Pomoni, M.~Taki and F.~Yagi,
	JHEP {\bf 1401}, 175 (2014)
  [arXiv:1310.3841 [hep-th]].
  
\bibitem{HKT}
  H.~Hayashi, H.~-C.~Kim and T.~Nishinaka,
	JHEP {\bf 1406}, 014 (2014)
  [arXiv:1310.3854 [hep-th]].

\bibitem{BZ} 
  O.~Bergman, G.~Zafrir,
	JHEP {\bf 1504}, 141 (2015)
  arXiv:1410.2806 [hep-th].

\bibitem{HTY}
  H.~Hayashi, Y.~Tachikawa and K.~Yonekura,
	JHEP {\bf 1502}, 089 (2015)
  [arXiv:1410.6868 [hep-th]].

\bibitem{OSTY2} 
  K.~Ohmori, H.~Shimizu, Y.~Tachikawa, and K.~Yonekura,
	JHEP {\bf 1512}, 131 (2015)
  [arXiv:1508.00915 [hep-th]].

\bibitem{KB}
  I.~Brunner, A.~Karch,
   Phys.\ Lett.\  B409:109-116 (1997)
  [arXiv:9705022 [hep-th]].

\bibitem{BZ1}
  O.~Bergman and G.~Zafrir,
	JHEP {\bf 1512}, 163 (2015)
	[arXiv:1507.03860 [hep-th]].

\bibitem{Zaf3} 
  G.~Zafrir,
  arXiv:1512.08114 [hep-th].
	
\bibitem{HKLTY1} 
  H.~Hayashi, S.~Kim, K.~Lee, M.~Taki, and F.~Yagi,
  arXiv:1512.08239 [hep-th].


	



 


\bibitem{Tachi1}
  Y.~Tachikawa, 
  JHEP {\bf 0907}, 067 (2009)
	[arXiv:0905.4074 [hep-th]].

\bibitem{CD1}
  O.~Chacaltana, J.~Distler,
  JHEP {\bf 1302}, 110 (2013)
  [arXiv:1106.5410 [hep-th]].

\bibitem{GRRY}
  A.~Gadde, L.~Rastelli, S.~S.~Razamat, and W.~Yan,
	Commun.\ Math.\ Phys.\ 252:359-391 (2004)
  [arXiv:1110.3740 [hep-th]].

\bibitem{GR}
  D.~Gaiotto, S.~S.~Razamat,
	JHEP {\bf 1205}, 145 (2012)
  [arXiv:1203.5517 [hep-th]].

\bibitem{GRR}
  D.~Gaiotto, L.~Rastelli, and S.~S.~Razamat,
	JHEP {\bf 1007}, 022 (2013)
  [arXiv:1207.3577 [hep-th]].

\bibitem{LPR}
  M.~Lemos, W.~Peelaers, and L.~Rastelli,
	JHEP {\bf 1405}, 120 (2014)
  [arXiv:1212.1271 [hep-th]].

\bibitem{CDT}
  O.~Chacaltana, J.~Distler, and A.~Trimm,
	JHEP {\bf 1509}, 007 (2015)
  [arXiv:1403.4604 [hep-th]].

\bibitem{Zaf2} 
  G.~Zafrir,
	JHEP {\bf 1512}, 157 (2015)
  arXiv:1509.02016 [hep-th].

\bibitem{Tachi2}
  Y.~Tachikawa,
  JHEP {\bf 1406}, 056 (2014)
  [arXiv:1402.4200 [hep-th]].

\bibitem{CDT1}
  O.~Chacaltana, J.~Distler, and A.~Trimm,
	JHEP {\bf 1504}, 173 (2015)
  [arXiv:1309.2299 [hep-th]].

\bibitem{HZ}
  A.~Hanany, A.~Zaffaroni,
  JHEP {\bf 9907}, 009 (1999)  
  [arXiv:9903242 [hep-th]].

\bibitem{HK}
  A.~Hanany, B.~Kol,
   JHEP {\bf 0006}, 013 (2000)
  [arXiv:0003025 [hep-th]].

\bibitem{Zaf1} 
  G.~Zafrir,
	JHEP {\bf 1507}, 087 (2015)
  arXiv:1503.08136 [hep-th].


	










	

\bibitem{BTX}
  F.~Benini, Y.~Tachikawa, and D.~Xie,
	JHEP {\bf 1009}, 063 (2010)
  [arXiv:1007.0992 [hep-th]].







\bibitem{KYW}
  A.~Kapustin, B.~Willett, and I.~Yaakov,
  [arXiv:1012.4021 [hep-th]].

\bibitem{CD}
  O.~Chacaltana, J.~Distler,
  JHEP {\bf 1011}, 099 (2010)
  [arXiv:1008.5203 [hep-th]].

\bibitem{Yak}
  I.~Yaakov,
	JHEP {\bf 1311}, 189 (2013)
  [arXiv:1303.2769 [hep-th]].

\end{thebibliography}
\end{document}